\documentclass[aps,prb,twocolumn]{revtex4-1}
\usepackage{graphicx}
\usepackage{ulem}
\usepackage{color}
 
\begin{document}

\title{Phase-resolved imaging of edge-mode spin waves using scanning transmission x-ray microscopy}


\author{C. Cheng\footnote{present address: School of Information Science and Electronic Engineering, Zhejiang University
38 Zheda Road, Hangzhou 310027, China}, W. Cao, and W. E. Bailey$^{*}$}
\affiliation{Materials Science and Engineering, Department of Applied Physics and Applied Mathematics, Columbia University, New York, NY 10027, USA}

\begin{abstract}
We have imaged the excitation of small-amplitude spin-wave eigenmodes, localized within $\sim$ 100 nm of the vertices of nanoscale Ni$_{81}$Fe$_{19}$ ellipses, using time-resolved scanning transmission x-ray microscopy (STXM) at 2 GHz and resolution of 70 nm.  Taking advantage of the buried-layer sensitivity of STXM, we find that the magnetization precession at the two vertices changes from predominantly in-phase to out-of-phase in samples with and without a conductive layer deposited over the ellipses.  As a plausible interpretation for the reversal in phase, we propose that unexpectedly strong Oersted fields are generated in the discontinuous overlayer, although effects of edge roughness cannot be fully excluded.  The results demonstrate the capabilities of STXM to image small-amplitude, GHz magnetization dynamics with the potential to map rf magnetic fields on the nanoscale. \end{abstract}
\maketitle

Spin-wave eigenmodes in magnetic nanostructures\cite{mathieuPRL98,JorzickPRL2002,parkPRL02} set the basis for the high-speed response of magnetoelectronic devices\cite{AlbertPRL2002,rippardPRL04, DemidovNMat2012}.  The eigenmodes comprise the spatial dependence of precessing (transverse) magnetization when the nanostructure is driven with ac fields $H_{rf}(\omega)$ near the mode frequency.  Techniques to image spin-wave eigenmodes in magnetic nanostructures are therefore of great interest.  Two techniques, near-field Brillouin light scattering ($\mu$-BLS) and magnetic resonance force microscopy (MRFM)\cite{mewesPRB06,obukhovPRL08}, have this capability, as they are able to resolve small-amplitude magnetization precession ($<$1$^{\circ}$ of rotation) at a spatial resolution (55 nm\cite{jerschAPL10} to 100 nm\cite{guoPRL13}, respectively) relevant for uniformly magnetized nanostructures ($\sim$ 500 nm width.)

A gap exists in the capabilities of these two techniques.  Imaging sensitive to the phase of the (complex) eigenmode is unknown for MRFM, attainable with some additional effort for $\mu$-BLS\cite{ulrichsPRB11}.  Imaging through conductive overlayers is natural for MRFM\cite{hamadehPRL13}, but a weaker point for $\mu-$BLS\cite{fohrPRL11}.  For device-relevant structures, both capabilities are important: the phase determines the local magnetoresistance response, and electrical leads are likely to be present if current is applied.   Time-resolved scanning-transmission x-ray microscopy (TR-STXM) offers both phase resolution, as a pump-probe technique, and the capability to image through thick conductive overlayers, due to the larger ($\mu$m) penetration depth of soft x-rays compared with the optical skin depth.  Moreover, it offers spatial resolution higher than near-field optical techniques, in principle to $\sim$ 15 nm\cite{attwoodNature05}, due to the short wavelength of x-rays.  This technique has been applied primarily to high-amplitude magnetization rotation, such as domain or vortex displacement in nonuniformly magnetized structures\cite{puzicJAP05,stollNature06}, as well as to high-amplitude motion below nanocontacts\cite{bonettiNComms15}, and only recently to small-angle dynamics\cite{ChengAPL2012,bonettiRSI15}.  

In this paper, we demonstrate TR-STXM as a technique to image spin-wave eigenmodes in nearly uniformly magnetized structures.  We image the lowest-frequency mode of 1000 nm $\times$ 500 nm $\times$ 20 nm Ni$_{81}$Fe$_{19}$ ellipses, the so-called "edge mode\cite{McMichaelJAP2005}," which has been imaged previously by scalar $\mu$-BLS\cite{jerschAPL10}.  Using the buried-layer and phase-sensitive detection of TR-STXM, we uncover surprising information on the excitation of the eigenmode.  Magnetization at the two edges precesses in the same direction (in phase, even symmetry) when no cap is added, but precesses in opposite directions (antiphase, odd symmetry) when the cap is added.  The changed behavior can be explained plausibly, but perhaps not uniquely, as the result of unexpectedly strong, odd-symmetry Oersted fields localized to the vertices, contributed by nanoscale topography in the cap.  The results suggest a technique to excite odd-symmetry eigenmodes in patterned structures, and seen another way, suggest the utility of STXM to detect rf magnetic fields\cite{baileyNComm13} at $\sim$ 100 nm resolution.

\begin{figure*}
  \includegraphics[width=2\columnwidth]{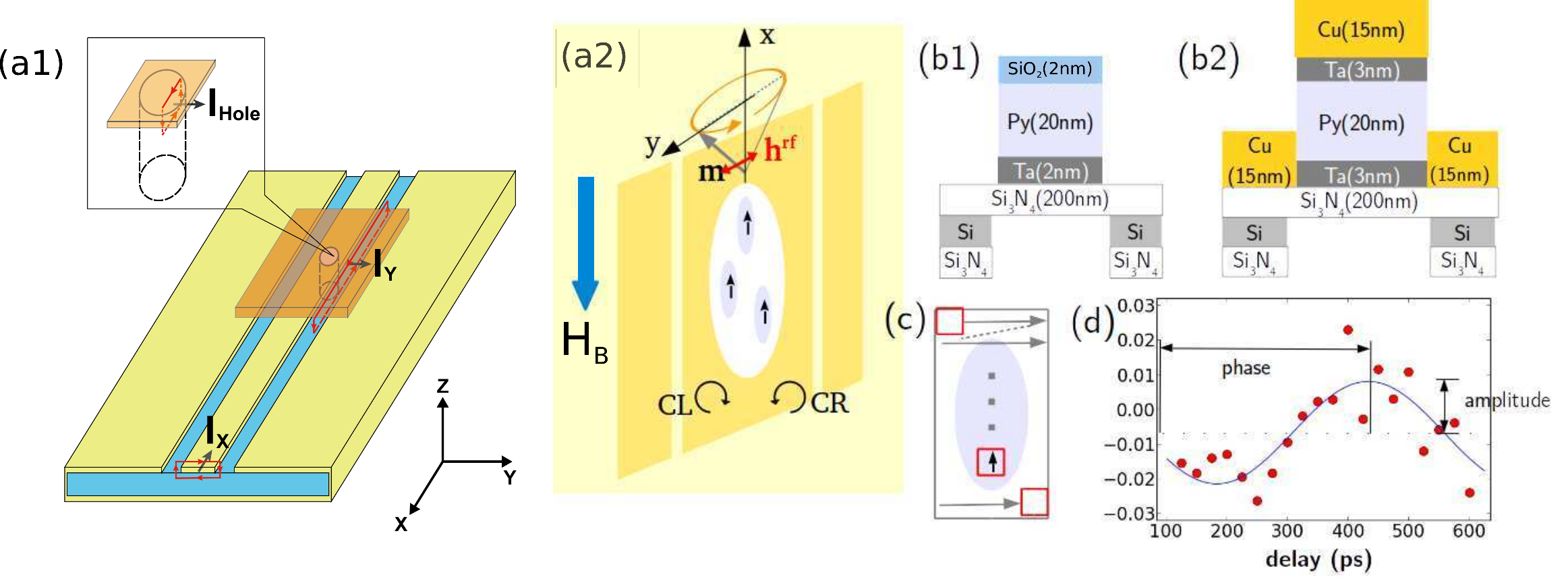}\caption{Experimental setup.  (a1)  Geometry of coplanar waveguide, with hole for x-ray transmission, and sample film plane, neglecting substrate, also as treated in finite-difference time-domain (FDTD) electromagnetic simulations\cite{oskooiCPC10}.  Amperean loops for calculations of the currents $i_x$ (along the CPW center conductor), $i_y$ (shorting signal to ground through the film), and $i_{hole}$ (in the vicinity of the Py ellipses) are illustrated.  (a2) Coplanar waveguide with circular hole, rotated 30$^{\circ}$ with respect to the circularly polarized incident beam direction $\mathbf{\sigma}=\pm\left(1/2\mathbf{y}+\sqrt{3}/2\mathbf{z}\right)$ (helicity CL, CR, as indicated.)  Ellipses are magnetized along the direction of the applied field $H_B\mathbf{x}$; $rf$ magnetic field $H_{rf}$ is on $\mathbf{y}$. (b)  Sample geometry, without (1) and with (2) the 15 nm conductive cap, looking along $\mathbf{x}$. (c)  Illustration of scan over ellipse, with $3\times 3$ averaging window. (d)  Sample TR-XMCD contrast and fit for XMCD phase and amplitude.}
  \label{fig0}
\end{figure*}

{\it Method:} TR-STXM measurements were carried out at the Canadian Light Source (CLS), soft x-ray spectromicroscopy beamline 10ID-1 (SM, x-ray spot size 40 nm).  The rf pump/x-ray probe technique has been described previously in Ref \cite{ChengAPL2012} demonstrated there for much larger elements (30 $\mu$m $\times$ 7 $\mu$m) in a multidomain state, and is illustrated in Figure \ref{fig0}.  X-ray pulses, here with $\hbar\omega=\textrm{707.9 eV}$ for x-ray magnetic circular dichroism (XMCD) contrast at the Fe L$_3$ edge, illuminate the sample with a repetition rate of 500 MHz and pulse width from the electron bunch width of $\sim$35 ps.  The bunch width determines the temporal resolution of a rising edge in TR-XMCD\cite{baileyPRB04}.  Phase-sensitive magnetization dynamics, and thus phase-sensitive spin-wave eigenmodes, are measured by locking the phase of a continuous-wave sinusoidal $rf$ magnetic field excitation (2GHz, +29 dBm), to the photon bunch clock (500 MHz), through upconversion with variable delay\cite{arenaRSI09}.  

\begin{figure}
  \includegraphics[width=\columnwidth]{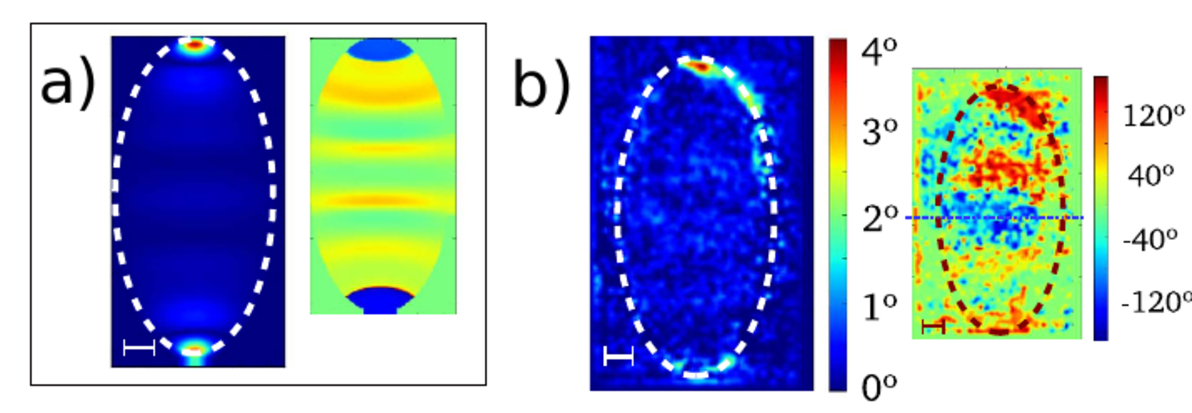}
  \caption{Predicted {\it(a)} and TR-STXM imaged {\it(b)} magnetization amplitude (blue background, {\it left}) and phase (green background, {\it right}), uncapped Ni$_{81}$Fe$_{19}$ ellipse (see Fig \ref{fig0}b1, driven at 2 GHz under zero bias (0 Oe). (c).  Scale bar 100 nm.}\label{fig1}
\end{figure}

For each image, the scanning step is fixed at 14.7 nm and averaged in a 3$\times$3-pixel square, yielding a spatial resolution of $\sim$ 70 nm with an x-ray spot size of 40 nm.  The 40 nm spot size is limited by the available zone plate optics at the beamline, and could in principle be reduced by a factor of two or more\cite{attwoodNature05}.  We recorded a series of 20 images for delay times spaced by 25 ns over the 500 ns period, increased to 30 images over 750 ns for the data in in Figure \ref{fig1}c.  Circular-right (CR) and circular-left (CL) images were subtracted from each other for each delay point.  Total data acquisition times per phase and amplitude image in \ref{fig1}c were roughly two hours, with each delay point scan pair requiring $\sim$3.5 min, and a dwell time per pixel of 2 ms.  The local window-averaged magnetic contrast at each pixel is then plotted as a function of the delay time, as illustrated in Figure \ref{fig0} c,d.  A sinusoidal fit to the local time-dependent magnetization $M(t)$ extracts the amplitude, phase and offset of the oscillation in contrast.  The XMCD amplitude is converted to magnetization angle (cone angle) through comparison with a sample fully saturated along the direction of the applied field.

Two separate Ni$_{81}$Fe$_{19}$ (Py) ellipse structures were imaged in the study, one with and one without a top Ta(3nm)/Cu(15nm) layer, as illustrated in Figure \ref{fig0} b1 and b2.  Ellipses with lateral dimensions of 1000 nm $\times$ 500 nm were defined by e-beam lithography and liftoff on commercially available Si$_3$N$_4$ membrane windows. The multilayer stacks were Ta (2 nm)/ Ni$_{81}$Fe$_{19}$ (20 nm)/ SiO$_2$ (2 nm) and Ta (3 nm)/ Ni$_{81}$Fe$_{19}$ (20 nm)/ Ta (3 nm) / Cu(15 nm), deposited by ultrahigh vacuum (UHV) sputtering at a base pressure of 2 $\times$ 10$^{-9}$ Torr\cite{chengRSI12}.  The films are expected to be nearly uniformly magnetized along the long axis of the ellipse: a saturating field of $\sim$ 500 Oe was applied along $\mathbf{x}$, using a permanent magnet, immediately prior to imaging, and the magnetization is expected from micromagnetic simulations to remain uniform along $\mathbf{x}$ at remanence.

In the experiment, the samples have been mounted to the surface of the coplanar waveguide (CPW), bridging the center conductor and ground shields, using conductive carbon tape.  The samples faced towards the CPW, and the carbon tape contacted the (conductive) front surface of the (Cu coated) film.  We therefore expect an ac, 2 GHz current path in $\mathbf{y}$, connecting signal to ground, across the thin-film for the sample covered with the Cu(15nm) layer; however, because the CPW is also shorted 1 mm above the hole, the contact is presumed, not verified.  The current is expected to be in the film plane, along the short axis of the ellipses ($\mathbf{y}$) with some spreading in the $\mathbf{x}$ direction.  No current path is possible for the patterned ellipses without the conductive overlayer.  A schematic for the film mounting on the CPW is shown in Figure \ref{fig2}a.

{\it Results:}  Fig. \ref{fig1} presents TR-STXM images of the magnetization amplitude and phase of single ellipses under $rf$ excitation at 2 GHz.  In the boxed panel, \ref{fig1}a, we compare the experimental images with the edge-mode response calculated through micromagnetic simulation at remanence ($H_{B}=0$) using the NMag software package\cite{FischbacherIEEE07}.  The calculation assumes a uniform transverse $rf$ field at 2 GHz, and shows the even 'edge' spin-wave eigenmode, concentrated within $\sim$ 100 nm of the vertices.  The even symmetry (magnetization of both vertices rotating in the same direction, in phase) is shown by the equal phase at the top and bottom, each lagging the $rf$ field by $\sim$90$^{\circ}$ (blue.)  This is the lowest-frequency spin-wave eigenmode of the ellipse.  Increasing the field in the direction of the magnetization ($H_B=\textrm{20,40 Oe}$) shifts all modes to higher frequency and decreases any relative spectral weight of the higher-order modes.  The next-higher-order mode has a 2 GHz resonance at fields opposite to the magnetization of $H_B\sim -\textrm{100 Oe}$.

The corresponding TR-STXM images for the uncapped sample, in Fig \ref{fig1}b, show a very similar localization of the transverse magnetization amplitude at the two ends (blue background) and mostly even phase response (green background.)  Here and in the following figure, we highlight that we have recorded is a {\it low-amplitude} response, with a maximum precessional cone angle of $\sim\textrm{6}^{\circ}$, calculated through the total variation in XMCD contrast.  In this sense, we demonstrate that spin-wave eigenmodes can be probed using TR-STXM as in the past using $\mu$-BLS\cite{ulrichsPRB11}.  

The central result of the paper is shown in Fig. \ref{fig2}b.  Here we have imaged the magnetization response for the ellipse with a Ta(3nm)/Cu(15nm) overlayer, as a function of bias field $H_B$ applied along the long axis of the ellipses ($\mathbf{H_B} = H_B\:\mathbf{x}$; see Fig \ref{fig0}a2.)  From the imaged amplitudes of the magnetization response, the edge mode is again clearly excited, although with mostly {\it odd} symmetry, seen in the (green background) insets of phase at 0 and 20 Oe.  

In the experimental images in Fig \ref{fig2}b, there is a large response of magnetization near the two vertices, with a $\sim$ 180$^{\circ}$ difference in phase.  At a higher bias field (20 Oe), the phase reverses at the two edges.   Very little magnetization precession is excited at lower (-10 Oe) and higher (40 Oe) fields.  This behavior is consistent with the existence of resonant field for the edge mode at 2 GHz at the low-field range of $\textrm{0 Oe}<H_B< \textrm{20 Oe}$, roughly consistent with the calculation. The odd-symmetry response is consistent with the results of a micromagnetic simulation assuming an odd-symmetry $rf$ driving field, $-H^{rf}_y(-x)=H^{rf}(x)$ (signum function in $x$), pictured in Fig \ref{fig2}a; the existence of degenerate even and odd-symmetry edge modes for the ellipse was reported previously by McMichael et al\cite{McMichaelJAP2005}.  

The excitation of an odd-symmetry spin wave eigenmode under the conductive cap, as observed by phase-sensitive TR-STXM in Fig \ref{fig2}b, can be explained most easily through the presence of an odd-symmetry external driving field.  That the two modes, symmetric and antisymmetric edge modes, are according to simulation equal in frequency, suggests that the magnetizations at the two edges are essentially uncoupled and respond only to the very local fields within $<$ 100 nm of the vertices of the ellipse.  Given that the $rf$ magnetic field is uniform on the scale of the CPW center-to-ground gap ($\sim$ 100$\mu$m) in the unloaded structure, it is initially very surprising that the local fields at the ellipse vertices, separated by 1 $\mu$m, could be opposite in sign.  

\begin{figure*}
  \includegraphics[width=2\columnwidth]{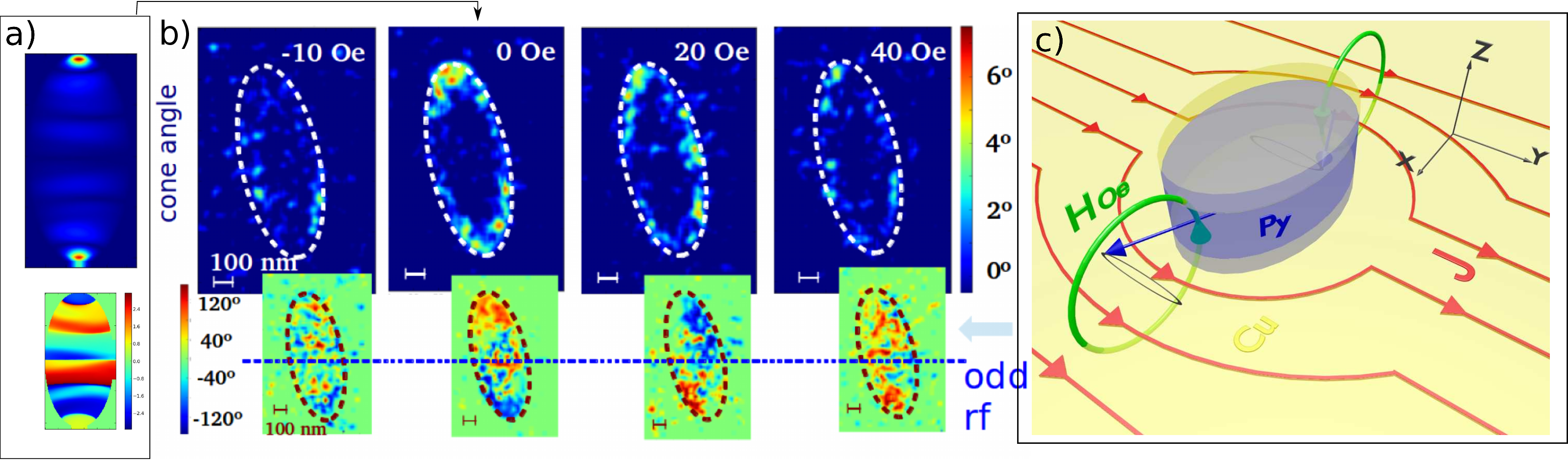}
  \caption{\label{fig2}(a): Simulation of amplitude and phase for the Ni$_{81}$Fe$_{19}$ ellipse, as in Fig \ref{fig1}a, but assuming odd-symmetry driving field $H_y(-x)=-H_y(x)$, resulting in odd-symmetry magnetization response.  (b)  As Fig \ref{fig1}b, for sample with Ta(3nm)/Cu(15nm) cap.  TR-STXM images of amplitude (top) and phase (bottom) for variable bias fields, showing odd phase response in $x$.  (c)  Illustration of proposed mechanism creating odd-symmetry $rf$ field: inhomogeneous in-plane current density $J_y(x,y)$ drives perpendicular $rf$ Oersted fields $\mathbf{H_{Oe}}= \pm H_{Oe}\mathbf{\hat{z}}$ and antisymmetric precession at the vertices of the ellipses.}\label{fig2}
\end{figure*}

The observation of even and odd-phase responses at resonance (near zero bias) is clearly above experimental noise.  The noise level considered within a $3\times 3$ pixel averaging window, under our imaging conditions, is roughly 0.5-1.0$^{\circ}$ of precession, corresponding to roughly 0.05 $\mu_B/\textrm{atom}$.  For the strongest signals, on resonance at $H=\textrm{0}$, the phase resolution is no better than $\pm\textrm{15}^{\circ}$.  The point phase resolution (at 70 nm resolution) is certainly poorer where the amplitude of precession is lower: in e.g. the center of the 0 Oe image in Fig \ref{fig1}c, away from the edge modes, there is mostly noise.  The noise in phase and amplitude improves if one averages with a larger window. 

{\it Discussion}:  We interpret the antiphase precession for the capped structure as the result of currents $J_y$ in the inhomogeneous conductive cap, as illustrated in Figure \ref{fig2}c.  We highlight here that the conductive path through the sample was not explicitly designed, verified, or controlled in the experiment except through the presence of the Cu layer, but would have arisen naturally as a result of the sample mounting, and appears to us to be the most plausible explanation for the observed phase behavior.  If a shorting current flows from the CPW signal to ground (in $\mathbf{y}$) across the cap, some net amount of the current, due to finite resistivity of the Cu film, will travel in $\mathbf{y}$ in the imaged area, which is offset from the exact center of the hole.  This transverse current $J_y$ will generate a perpendicular magnetic field $H_z$ at the vertices of the ellipses in the structures investigated.  The Ta(3nm)/Py(20nm)/Ta(3nm) ellipses were patterened first by e-beam lithography, and the Cu(15nm) layer was deposited over them in a second step, as shown in Figure \ref{fig0}b2.  The field region has uniformly thick Cu(15nm), but there is therefore a relatively large topographic step, greater than the Cu thickness, at the ellipse edge.  Direct contact of uniformly thick Cu is not likely.  

The magnetic fields generated near the ellipses could be modeled most simply as those arising from a current-carrying plate with elliptical holes.  Far from the holes, the plate carries a uniform current $J$; in the holes, there is no current, and the current crowds around the edges of the holes.  Even if there is excellent conductive contact between the planar Cu and the Py ellipse, which is not expected, the Py ellipse has a much higher bulk resistivity (by a factor of 3-4), thus the in-plane current $J_y$ will crowd around the vertices of the ellipse.  The magnetostatic fields for this simple model have been calculated in detail in Ref \cite{jinIEEE09}; we can consider the limiting case of a round hole with radius $a$.  Here, there will be a perpendicular field at the ellipse edges, $H_z(x=\pm a)=\mp J_0 a$, as illustrated in Fig \ref{fig2}c.  This field has the correct odd symmetry to explain the antiphase precession at the vertices when the conductive cap is present.  Moreover, we expect this field to be roughly large enough to excite the dynamics observed, based on simulations of the current density $J_y$ and micromagnetics of the ellipses.

 We have estimated $J_y$ in the Cu(15nm) overlayer using finite-difference time-domain electromagnetic simulations with public software (MEEP\cite{oskooiCPC10}).  Our three-dimensional geometry, shown in Figure \ref{fig0}a1, includes the CPW, with hole for x-ray transmission, and Cu(15nm) film, neglecting the Si/Si$_3$N$_4$ substrate.  The film is in conductive contact with the ground and signal lines of the CPW, shorting the primary current.  We find the ratio of current through the CPW center conductor $i_x$ and the sample region $i_{hole}$ using Amperean loops $\oint \mathbf{H}\cdot d\mathbf{l} = i$, as drawn, under 2 GHz excitation.  We hold the film conductance $R_{sq}^{-1}=t_{Cu}/\rho_{Cu}$ to the experimental value while increasing $t_{Cu}$ and $\rho_{Cu}$ proportionally to make meshing possible.\cite{bailleulAPL13}.  Most of the current is sustained near the leading ($+x$) edge of the film, but some current density $J_y$ exists over the full film; we estimate $i_{hole}/i_x \simeq \textrm{1/400}$.  This leads to a current density in the Cu(15nm) overlayer of $J_y \sim \textrm{2}\times\textrm{10}^{8}\textrm{A/m}^2$ at 1 W.

According to the micromagnetic simulations, the edge modes have a precessional ellipticity on resonance at 2 GHz of $\eta=m_{y}/(-i m_z) \sim \textrm{6}$.  This ellipticity is reduced from the infinite-film value for Py at 2 GHz of $\sim$14, due to the more nearly spherical volume of precession in the edge modes.  Perpendicular $rf$ magnetic fields $H_z$ larger than in-plane $rf$ magnetic fields $H_y$ by a factor of $\eta$ are required to excite similar precessional amplitudes, so 2 GHz fields $H_z \sim \textrm{240 A/m}$ at the ellipse vertices are required for the observed dynamics.  These odd-symmetry fields would be produced at the ellipse vertices by current densities $J_y\sim H_z/a\sim {6}\times\textrm{10}^{8}\textrm{A/m}^2$ for $a=\textrm{375 nm}$, in order-of-magnitude agreement with the current densities expected from the electromagnetic simulation.

Another plausible origin for the antiphase precession is nonuniformity of the tips.  If the resonance frequencies for the top and bottom edges differ, due to e.g. edge roughness, one edge could be excited above its resonance and the other below its resonance at zero field.  For Py at 2 GHz, in order for there to be a 150$^\circ$ phase difference, the resonances need to differ by more than twice the half-power linewidth, a minimum of $\sim$ 20 Oe, split to high and low field.  While some nondegeneracy is certainly possible, we find that it does not agree with the available images as the mechanism producing the antiphase precession.  First, the amplitudes of edge-mode precession at 0 Oe are higher than those at 20 Oe by more than a factor of two for both top and bottom edges, implying that {\it both} resonances are closer to 0 Oe than 20 Oe.  If one edge had a higher-field resonance, its precessional amplitude should increase.  Second, in the experiment, the phases change in opposite directions from $H_B=0$ to $H_B=\textrm{20 Oe}$: $\sim-90^{\circ}$ to $\sim0^{\circ}$ at the top and $\sim+90^{\circ}$ to $\sim0^{\circ}$ at the bottom.  For split resonances, they should both decrease, the low-field resonance by a small amount and the high-field resonance by a larger one.  Some role of edge roughness is certainly possible, given that the symmetry of the images is not perfectly odd or even, but it does not appear to be the primary driver of antiphase precession in the Cu-capped samples.

{\it Conclusion:}  We comment on some prospects for the technique.  The noise level under our imaging conditions is roughly 0.5-1.0$^{\circ}$ of precession, corresponding to roughly 0.05 $\mu_B/\textrm{atom}$, with a best phase resolution of $\pm\textrm{15}^\circ$.  It is not feasible to achieve another order of magnitude of sensitivity by averaging many multiples of the 110-minute acquisition period per image.  Higher resolutions may be possible with a synchronous measurement technique.  This approach increases the sensitivity and resolution of broad-beam TR-XMCD by nearly two orders of magnitude\cite{arenaRSI09}, but requires a different implementation in STXM\cite{chengJAP12} which is cumbersome to set up in short beamtime allocations.  Recently, a remarkably high detection sensitivity of $\sim\textrm{6}\times\textrm{10}^{-5}\mu_B/\textrm{at}$ for transient induced moments in Cu(28nm) was reported using synchronous detection of current modulation in STXM\cite{kukrejaPRL15}, so higher-resolution imaging of less intense eigenmodes, in thinner films, may yet prove possible.

{\it Summary: }  We have demonstrated time-resolved scanning transmission x-ray microscopy (TR-STXM) as a technique to image the small-amplitude spin-wave eigenmodes of nearly uniformly magnetized magnetic nanostructures.  The spin-wave edge-modes of 1000 nm $\times$ 500 nm $\times$ 20 nm Ni$_{81}$Fe$_{19}$ ellipses show a resonant response as a function of applied field at 2 GHz.  The advantages of the technique, specifically spatial resolution, phase-sensitivity, and the ability to measure through conductive caps, have allowed us to visualize a dramatic effect of the cap layer on local fields, interpreted as the role of nanoscale topography on Oersted fields.  The results and technique demonstrated are relevant for the interpretation of $rf$ magnetization dynamics experiments on magnetic nanostructures.  

We thank Jian Wang (CLS) for technical assistance at the beamline, and acknowledge NSF-ECCS-0925829 and NSF-DMR-1411160 for support.

\end{document}